\documentclass[prl,twocolumn,superscriptaddress]{revtex4-2}
\usepackage{graphicx}
\usepackage{amsmath}
\usepackage{amssymb}
\usepackage{bm}
\usepackage{color}
\usepackage{hyperref}
\usepackage{multirow}
\usepackage{braket}
\usepackage{epstopdf}
\usepackage{physics}
\usepackage{booktabs}
\usepackage{graphicx}
\usepackage{array}[=2016-10-06]

\makeatletter
\def\maketitle{
  \@author@finish
  \title@column\titleblock@produce
  \suppressfloats[t]}
\makeatother

\newcommand{\ii}{\mathrm i}
\newcommand{\vF}{v_{\!F}}
\newcommand{\Ds}{\Delta_s}

\newcommand{\twoFVsWeightColumnWidth}{0.32\columnwidth}
\newcommand{\twoFVsContourColumnWidth}{0.37\columnwidth}
\newcommand{\twoFVsPeakColumnWidth}{0.20\columnwidth}
\newcommand{\threeFVsWeightColumnWidth}{0.32\columnwidth}
\newcommand{\threeFVsGeometryColumnWidth}{0.44\columnwidth}
\newcommand{\threeFVsPeakColumnWidth}{0.23\columnwidth}

\newcommand{\be}{\begin{equation}}
\newcommand{\ee}{\end{equation}}
\newcommand{\bea}{\begin{eqnarray}}
\newcommand{\eea}{\end{eqnarray}}

\begin{document}
\title{Majorana Zero Modes in Fractional Vortices}
\author{Lecheng Fang}
\author{Haotian Li}
\author{Haijiao Ji}
\affiliation{Tsung-Dao Lee Institute, Shanghai Jiao Tong University, Shanghai 201210, China}
\affiliation{School of Physics and Astronomy, Shanghai Jiao Tong University, Shanghai 200240, China}
\author{Quanxin Hu}
\email{qxhu@xmu.edu.cn}
\affiliation{Department of Physics, College of Physical Science and Technology, Xiamen University, Xiamen 361005, China}
\author{Hong Ding}
\email{dingh@sjtu.edu.cn}
\author{Noah F. Q. Yuan}
\email{yuanfanqi@sjtu.edu.cn}
\affiliation{Tsung-Dao Lee Institute, Shanghai Jiao Tong University, Shanghai 201210, China}
\affiliation{School of Physics and Astronomy, Shanghai Jiao Tong University, Shanghai 200240, China}

\begin{abstract}
Majorana zero modes (MZMs) are promising for topological quantum computation, with vortices in topological superconductors serving as a leading platform. Recent experiments show that integer Abrikosov vortices in iron-based superconductors can split into fractional vortices (FVs) carrying fractional flux quanta. We show that due to topological protection, MZMs hosted by FVs are spatially unfolded over the FV cores, while fermionic bound states remain localized at individual FV cores. This yields zero-bias conductance peaks in both local tunneling and nonlocal crossed Andreev reflection, offering an alternative MZM signature.
\end{abstract}
\maketitle

\textit{\textcolor{blue}{Introduction.}}---
Majorana zero modes (MZMs) are zero-energy bound states in topological superconductors (TSCs) and are central to fault-tolerant topological quantum computation \cite{Kitaev2001,ReadGreen2000,Lutchyn2010}. They appear at the ends of one-dimensional TSCs and in vortices of two-dimensional TSCs, where their non-Abelian statistics enable topological braiding \cite{Ivanov2001,Nayak2008,Alicea2012,Beenakker2013,Teo2010}. The Fu-Kane model \cite{FuKane2008} binds an MZM to an integer Abrikosov vortex on a proximitized topological-insulator surface, for which iron-based superconductors are intrinsic candidates where topological surface states coexist with bulk superconductivity \cite{FeSCtheory0,FeSCtheory1,FeSCtheory2,FeSCtheory3}.

Recently, \textit{fractional vortices} (FVs) carrying fractions of one flux quantum were directly observed in iron-based superconductors \cite{fractional2026,Zhou2024VortexSplitting,MultibandSC}, further expanding the defect landscape that can support MZMs. Unlike an integer Abrikosov vortex, FVs can form composite defects with internal structures such as skyrmions \cite{Babaev0,Babaev1,Babaev2,Lindner2012,CP2sky1,CP2sky2}. This naturally raises a question: how do MZMs organize in FVs?

On the other hand, the standard detection method of MZMs is local tunneling conductance, where a zero-bias conductance peak (ZBCP) quantized at \(2e^2/h\) signals an MZM \cite{Law1,Law2,CWJ1,CWJ2}. The ZBCPs have been reported in iron-based superconductors at cores of Abrikosov vortices \cite{FeSCexp2018,FeSCexp2019,Liu2018,Liu2020,FeSCexp4,FeSCexp5,FeSCexp6,FeSCexp7}. 
However, a ZBCP alone can be mimicked by trivial Caroli-de Gennes-Matricon (CdGM) bound states \cite{CdGM0,Hess1989,Hess1990} and impurity resonances \cite{JLiu2012}. 

This naturally raises another question: how to distinguish MZMs from trivial bound states, especially in FVs?

In this work, we address these two questions by theoretically studying generic vortex bound states of FVs in TSCs. We show that, due to the topological protection, MZMs in FVs are spatially unfolded over the FV cores, whose size can be controlled by the FV-core separations [Fig.~\ref{fig1}(a-c)]. In contrast, fermionic CdGM bound states remain localized at individual cores [Fig.~\ref{fig1}(d-f)]. As a result, the ZBCP appears not only locally but also in nonlocal crossed Andreev reflection (CAR) at long distances, as long as they are smaller than the MZM size (Fig.~\ref{fig2}). The ZBCP of nonlocal CAR is not expected from fermionic CdGM states or impurity resonances, providing an alternative experimental signature of MZMs.

In the following, we first extend the Fu-Kane model from Abrikosov vortices to FVs. Then we derive the MZM wavefunctions and analyze the spatial and spin distributions. 
We find that the MZM is spatially unfolded compared with the single Abrikosov vortex case, with a helical spin texture along the line joining the FV cores. 
We thus study nonlocal crossed Andreev reflection enhanced by this unfolded spatial configuration. Finally, we discuss the effect of bulk states and examine the stability of surface MZMs against them.

\begin{figure}
    \centering
    \includegraphics[width=1.0\linewidth]{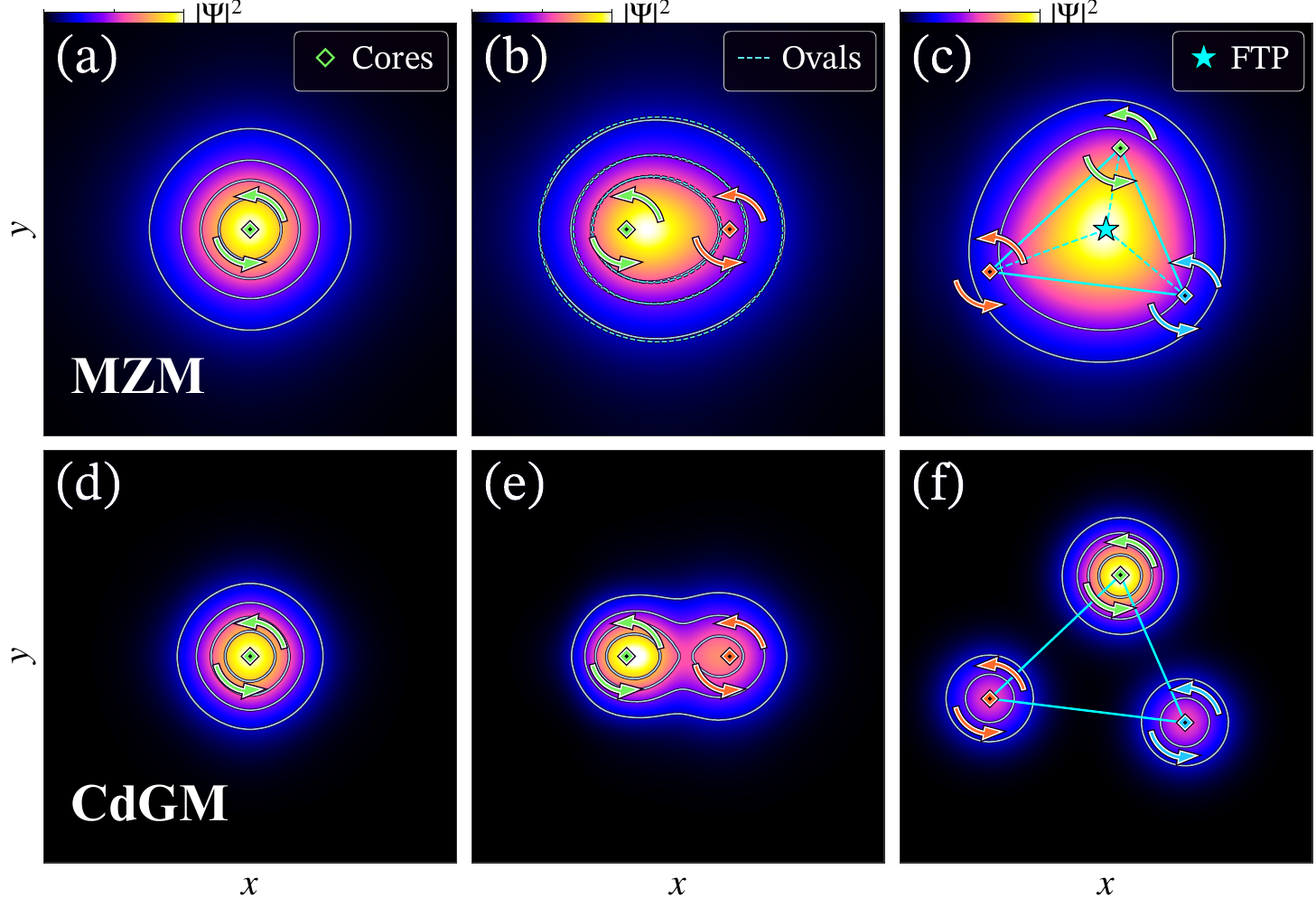}
    \caption{Probability density $|\Psi|^2$ of (a-c) Majorana zero modes (MZMs) and (d-f) fermionic Caroli--de Gennes--Matricon (CdGM) bound states under vortex-core fractionalization from Abrikosov vortices (a,d) to fractional vortices (b,c,e,f). Peaks of $|\Psi|^2$ in (a-c) are at the generalized Fermat--Torricelli points (FTPs), whereas those in (d-f) are at individual vortex cores. We used the model of Eq. (\ref{eq_1}) with parameters and details given in the Supplementary Material.}
    \label{fig1}
\end{figure}

\textit{\textcolor{blue}{Fu-Kane-type model with fractional vortices.}}---
We consider a three-dimensional (3D) topological material with multiple
orbitals and \(s\)-wave pairing potentials \(\Delta_{\alpha}\) labeled by
orbital index \(\alpha\). Vortex lines are introduced into
\(\Delta_{\alpha}(\bm r)\), stretching between the top and bottom
surfaces. We neglect the external magnetic field that generates the
vortices and work in the extreme type-II limit.

When all vortex cores of \(\Delta_{\alpha}(\bm r)\) coincide, MZMs are
well known to appear at the ends of the Abrikosov vortex lines
\cite{FuKane2008,Hosur2011}, as in Kitaev chains \cite{Kitaev2001}. We
now separate the vortex cores of different orbitals and investigate MZMs
in such FVs, which can be viewed as an analogue of multi-orbital Kitaev
chains for FV lines. MZMs still appear at the ends of the FV lines,
but now carry orbital weights and orbital distributions. Because the
orbitals are spatially separated, this orbital distribution is
manifest not only in orbital space but also in real space.

Without loss of generality, we focus on the topological surface states (TSSs) \(\psi_{\uparrow,\downarrow}\) on the top surface, whose effective Bogoliubov--de Gennes (BdG) Hamiltonian is %Fu-Kane-type
\begin{equation}\label{eq_surf}
H=
\begin{pmatrix}
v_{F}(\bm\sigma\times\bm k)\cdot\hat{\bm z}-\mu & \Delta(\bm r)\\
\Delta^{*}(\bm r) & \mu-v_{F}(\bm\sigma\times\bm k)\cdot\hat{\bm z}
\end{pmatrix}
\end{equation}
on the Nambu basis \((\psi_{\uparrow}^{\dagger},\psi_{\downarrow}^{\dagger},\psi_{\downarrow},-\psi_{\uparrow})\), with momentum \(\bm k=(k_{x},k_{y})\), Fermi velocity \(v_F\), normal vector \(\hat{\bm z}\), chemical potential \(\mu\), and spin Pauli matrices \(\bm\sigma\).

The TSS wavefunctions \(\psi_{\uparrow,\downarrow}\) are specific linear superpositions of the multi-orbitals, and \(\Delta(\bm r)\) is the TSS pairing obtained by projecting the pairing functions \(\Delta_{\alpha}(\bm r)\) onto the manifold spanned by \(\{\psi_{\uparrow},\psi_{\downarrow}\}\), which is therefore a linear superposition of the vortex pairing functions. Hence, we adopt the TSS pairing potential ansatz
\begin{equation}
\Delta(\bm r)=\sum_{\alpha}f_{\alpha}(\rho_{\alpha})e^{i\theta_{\alpha}},
\end{equation}
with \(\rho_{\alpha}=|\bm r-\bm r_{\alpha}|\) and \(\theta_{\alpha}=\arg(\bm r-\bm r_{\alpha})\), up to an overall U(1) phase. Here, \(\bm r_{\alpha}\) is the vortex-core site of orbital \(\alpha\), and the pairing amplitude \(f_{\alpha}\geq 0\) vanishes at the FV core, \(f_{\alpha}(0)=0\), and becomes asymptotically uniform, \(f'_{\alpha}(\infty)=0\). 
With \(n\geq 2\) distinct FV-core sites \(\{\bm r_{\alpha}\}\), the FVs form \(\mathbb{C}P^{n-1}\) skyrmions \cite{Babaev0,Babaev1,Babaev2,Lindner2012,CP2sky1,CP2sky2}. 

We look for an MZM solution \(H\Psi_0=0\) to the surface effective Hamiltonian Eq.~(\ref{eq_surf}).
The existence of MZMs in the TSS pairing potential \(\Delta(\bm r)\) is guaranteed by the index theorem, since the phase of \(\Delta(\bm r)\) winds by \(2\pi\) along a contour enclosing all FV cores. We therefore expect the spatial extent of the MZMs to encompass all FV cores, so as to reveal the integer phase winding. 

Examples of the spatial distribution of MZM density \(\Psi_0^{\dagger}\Psi_0\) can be numerically worked out as shown in Fig. \ref{fig1}(a-c), which are spatially unfolded over the FV cores, in contrast to fermionic CdGM bound states which are localized at individual cores as shown in Fig. \ref{fig1}(d-f). Details of numerical simulations can be found in the Supplementary Material (SM). In the following we will present the analytical expression of MZM wavefunction.

\textit{\textcolor{blue}{Majorana wavefunction.}}---
We first focus on the TSS Dirac point \(\mu=0\) with chiral symmetry \(\{\sigma_z\tau_z,H\}=0\), where Pauli matrix $\tau_z$ acts in particle-hole space. Then MZM is also a chiral eigenstate %$\sigma_z\tau_z\Psi_0=\Psi_0$,
\begin{equation}\label{eq_MZM}
\Psi_0=\mathcal{N}\chi e^{-W/v_F},
\quad
W(\bm r)=\sum_{\alpha}\int_0^{\rho_{\alpha}}f_{\alpha}(\rho)\,d\rho,
\end{equation}
with normalization constant \(\mathcal{N}\) and \(\chi=(1,0,0,-1)^{\rm T}\).

The spatial distribution of MZM density \(\Psi_0^{\dagger}\Psi_0\) is hence characterized by function $W(\bm r)$.
At sufficiently large distances from the FV cores, the pairing potential amplitudes saturate to uniform pairings \(\Delta_{\alpha}^{\infty}\equiv f_{\alpha}(\infty)\) and \(W=\sum_{\alpha}\Delta_{\alpha}^{\infty}\rho_{\alpha}\). Thus, the MZM density contours away from the vortex cores become generalized Cartesian ovals \(\sum_{\alpha}w_{\alpha}\rho_{\alpha}=C\), where \(w_{\alpha}=|\Delta_{\alpha}^{\infty}|/|\Delta_{\infty}|\) is the orbital weight measured by the surface pairings, with \(|\Delta_{\infty}|^2=\sum_{\alpha}|\Delta_{\alpha}^{\infty}|^2\). The MZM density peak is at the point where \(\sum_{\alpha}w_{\alpha}\rho_{\alpha}\) is minimized, which is by definition the generalized Fermat--Torricelli point (FTP) of the FV-core sites \(\{\bm r_{\alpha}\}\) with weights \(\{w_{\alpha}\}\). The spatial extent of the MZM from the FTP is controlled by the FV-core separations \(\{|\bm r_{\alpha}-\bm r_{\beta}|\}\), the orbital weights $\{w_{\alpha}\}$ and the coherence length \(\xi\equiv v_F/|\Delta_{\infty}|\) as expected.

When \(n=1\), the MZM density contours are circles with the FTP at the vortex core, as shown in Fig.~\ref{fig1}(a). Besides Abrikosov vortices, this case can also apply to regions sufficiently far from the FV cores, \(|\bm r|\gg|\bm r_{\alpha}-\bm r_{\beta}|\).

When \(n=2\), the MZM density contours are Cartesian ovals \(w_1\rho_1+w_2\rho_2=C\). When \(w_1\neq w_2\), without loss of generality we assume \(w_1>w_2\), then the MZM density peak, the FTP, is at focus \(\bm r_1\), as shown in Fig.~\ref{fig1}(b). When \(w_1=w_2\), the contours become confocal ellipses \(\rho_1+\rho_2=2C\) with two foci at \(\bm r_{1,2}\), and the MZM density peak is not a single point but a continuous line segment joining \(\bm r_1\) and \(\bm r_2\), which may be called the generalized Fermat--Torricelli line (FTL), as shown in Fig.~\ref{fig2}(a). When \(w_2=0\), the surface states are fully polarized to orbital 1, and the MZM density contours are circles centered at \(\bm r_1\). When \(\bm r_1\) is so far from \(\bm r_2\) that the FV at core \(\bm r_1\) can be treated as a single FV, the MZM density contours are ellipses with one focus at \(\bm r_1\) and eccentricity \(w_2/w_1<1\). These examples are elaborated in the SM.

When \(n=3\), the FTP can be an interior point of the triangle formed by the three FV cores [Fig.~\ref{fig1}(c)], one of the FV cores, or can evolve into an FTL, i.e., the line segment joining the two nearest FV cores, as elaborated in the SM.

Next, based on the Fu-Kane mechanism \cite{FuKane2008}, we argue that our results for unfolded MZMs can also apply to \(\mu\neq 0\), which is consistent with our numerical simulations.

When the FTP (or FTL) is not at any FV core, the TSS pairing vanishes at the FTP (or FTL) due to destructive interference of the pairings from different FVs. In this case, the TSS pairing potential can be treated as a single vortex with its core at the FTP (or FTL), and the MZM is topologically protected by the integer phase winding of the TSS pairing around the FTP (or FTL).

When the FTP is at one of the FV cores (say orbital \(\alpha\)), orbital \(\alpha\) dominates the surface pairing and supports MZMs through the Fu--Kane mechanism, with modifications from the other orbitals. In this case there is no zero point of the TSS pairing, since the essence of the Fu--Kane mechanism lies in the integer winding of the superconducting phase, rather than in the spatial nodes of the order parameter.

Numerical results for the MZM density at \(\mu\neq 0\), including FTPs and FTLs, can be found in the SM.

The local density of states (LDOS) at zero energy is proportional to the MZM density. Thus, in experiments, by measuring the LDOS (e.g., via tunneling spectroscopy) in FVs, one could distinguish MZMs [Fig.~\ref{fig1}(a-c)] from fermionic CdGM bound states [Fig.~\ref{fig1}(d-f)].

Unlike LDOS, the spin configurations of MZMs depend sensitively on $\mu$.
Notice that at $\mu=0$, MZM in Eq. (\ref{eq_MZM}) is composed of spin up components ($\psi_{\uparrow}$ and $\psi_{\uparrow}^{\dagger}$) only due to chiral symmetry.
When $\mu\neq 0$, chiral symmetry is broken, and spin down components will develop.
The spin texture of MZM is the real-space fingerprint of the topological phase winding of FVs, transmitted to the spin sector by strong spin–orbit coupling of TSS Dirac Hamiltonian, and can be detected by the spin-resolved LDOS measured by a spin-polarized lead at zero energy.

\textit{\textcolor{blue}{Enhanced crossed Andreev reflections.}}---
Besides the extended spatial distribution of LDOS and local spin configurations, nonlocal transports such as crossed Andreev reflections (CARs) could also be able to distinguish MZMs from fermionic CdGM bound states.

A CAR is a nonlocal quantum transport process in a normal–superconductor–normal (NSN) junction \cite{CAR0,CAR1,Law1,CAR3,CAR4} where an electron entering one normal lead is reflected as a hole at a spatially separated lead, as shown in Fig. \ref{fig2}(a). 
The nonlocality of CAR usually requires phase coherence across the superconductor over distances comparable to the coherence length.
In the unfolded MZMs, the phase coherence is controlled not only by the coherence length, but also the FV-core separations. Thus we expect CAR over distances beyond the coherence length.

The CAR is amplified in the case of FTL formed by two FVs as shown in Fig. \ref{fig2}(a), where two leads are inside the FTL with separation $d$, and FV-core separation $a$. 
Since the pairing correlation $\langle\psi_{\sigma}(\bm r)\psi_{\sigma'}(\bm r')\rangle$ is spatially constant as long as $\bm r,\bm r'\in$ FTL, it can be found that CAR conductance does not significantly drop when $d$ increases, as long as both leads are inside the FTL. 
In our setup, one lead is fixed at the center of FTL [Fig. \ref{fig2}(a)], thus the significant drop is at $d\sim a/2$ [red line in Fig. \ref{fig2}(b)]. For comparison, we also plot CAR in the single Abrikosov vortex case $a=0$, whose significant drop is at $d\sim\xi$ (blue line).

Our unfolded MZMs especially FTLs offer good platforms for CARs between FVs at distances beyond coherence length.
In experiments of iron-based superconductors, FVs are usually found collinear \cite{fractional2026}, where enhanced CARs are expected. To carry out the measurements, on-chip multi-tip platform \cite{Maarten2020} and carbon nanotube tips \cite{CNTprobe1,CNTprobe2,CNTprobe3} may be employed.

\begin{figure}
    \centering
    \includegraphics[width=1.0\linewidth]{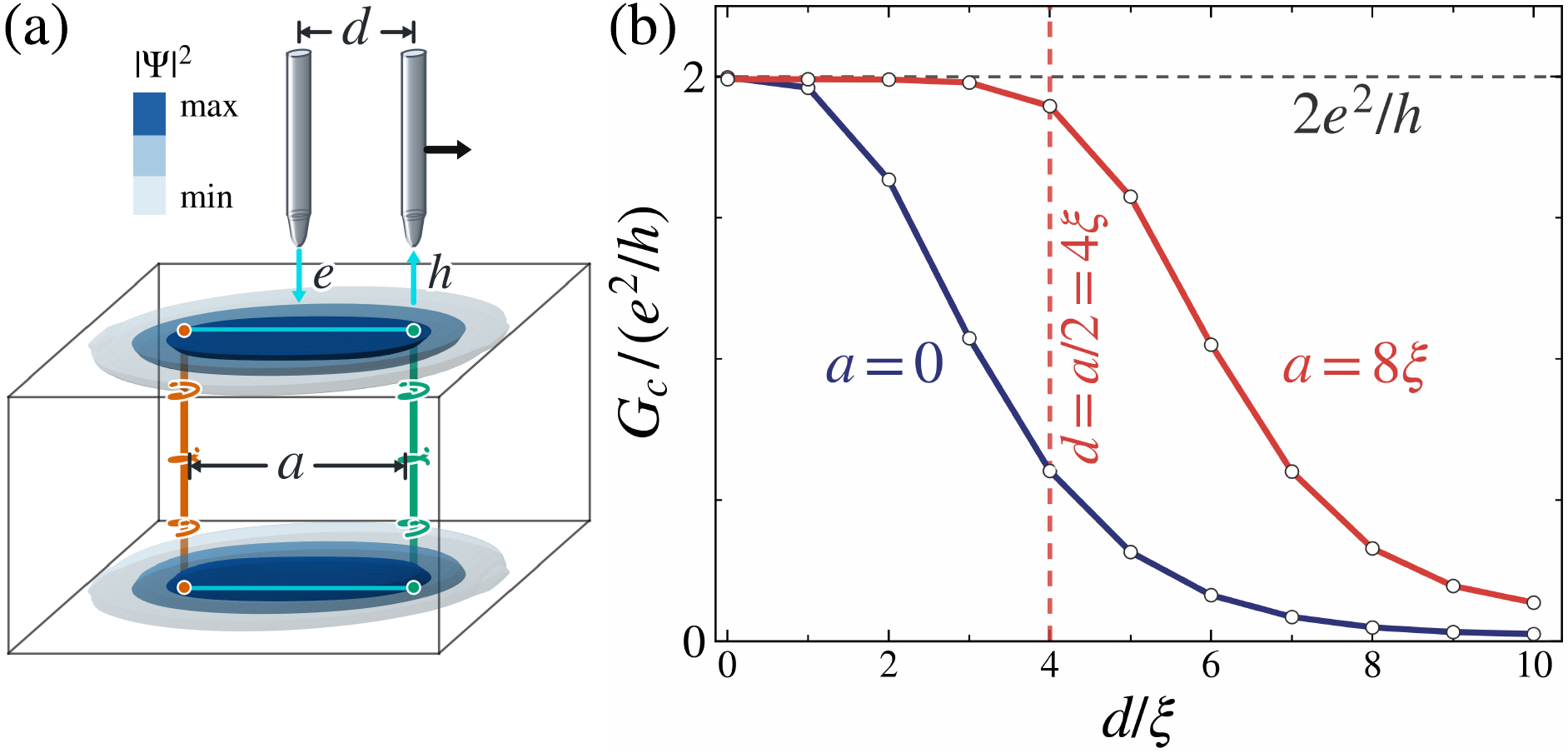}
    \caption{(a) Setup and (b) tunneling conductance of crossed Andreev reflections by Majorana zero modes in collinear fractional vortices. The blue line in (a) denotes the Fermat–Torricelli Line (FTL). We used the lattice model with parameters \(t=0.4\), \(M=2.0\), \(m_0=-1.0\). The lead hopping is $t_L = 0.8$. %The hopping between the lead and the surface is $t_c = t/2=0.2$.
    The lead–surface hopping is $t_c = 0.2$.}
    \label{fig2}
\end{figure}

\textit{\textcolor{blue}{Bulk state effects.}}---
In the previous sessions we have considered MZMs purely from topological surface states.
When bulk states are involved, MZMs start to extend more to the bulk and eventually penetrate through the vortex lines, enabling the hybridization between top and bottom surfaces.
As a result, the vortex phase transitions (VPTs) between topological and trivial phases would happen when bulk states are sufficiently involved \cite{Hosur2011}.
To include bulk states, the full BdG Hamiltonian of the 3D superconductor is
\begin{equation}\label{eq_1}
  \mathcal{H}=\begin{pmatrix}
H_0(\bm k) & \hat\Delta(\bm r)\\
\hat\Delta^{\dagger}(\bm r) & -H_0(\bm k)
\end{pmatrix},
\end{equation}
on the Nambu basis $(\bm c_{\uparrow}^{\dagger},\bm c_{\downarrow}^{\dagger},\bm c_{\downarrow},-\bm c_{\uparrow})$, where $\bm c_{\sigma}=\{c_{\sigma\alpha}\}$ is assumed to have spin $\sigma=\uparrow,\downarrow$ and $\alpha=1,\dots,N$ orbital components, and $\bm k=-i\nabla$ is the momentum operator.

Due to the bulk-edge correspondence, the existence of MZMs on the open surfaces is determined by the topology of bulk states under the periodic boundary condition (PBC) along the vortex line direction. Under the $z$-axis PBC, the bulk Hamiltonian can be labeled by the $z$-axis momentum $k_z$ as $\mathcal{H}=\mathcal{H}(k_z)$, which belongs to class D in one dimension \cite{class1,class2}, with the topological invariant $\nu\in\mathbb{Z}_2=\{0,1\}$ defined by
\be
(-1)^{\nu}={\rm sgn}\prod_{k_z=0,\pi}{\rm Pf}[U_P \mathcal{H}(k_z)],
\ee
where Pf denotes Pfaffian, and $U_P$ is the unitary part of the particle-hole symmetry operator.

At weak pairings, the VPT of an Abrikosov vortex is expected to be governed by properties of the bulk state Fermi surface (FS).
The VPTs of FVs and skyrmions in general depend on not only properties of bulk states (e.g. Fermi energy and Berry phase of FS), but also the internal structures of the skyrmions (e.g. orbital weights, FV-core separations and orientations).

For demonstration, we consider the lattice model where one Abrikosov vortex can be split into two FVs.
In this case, the 4$\times$4 TI Hamiltonian is expanded in terms of one identity matrix and four Dirac Gamma matrices $\{\Gamma_i,\Gamma_j\}=2\delta_{ij}$ $(i,j=0,1,2,3)$ as
$H_0(\bm k)=\varepsilon(\bm k)+\bm d(\bm k)\cdot\bm\Gamma+m(\bm k)\Gamma_0-\mu$,
with $\varepsilon(\bm k)=-2q(\cos k_x+\cos k_y+\cos k_z)$, $\bm d(\bm k)=2t(\sin k_x,\sin k_y,\sin k_z)$ and mass term $m(\bm k)=M+m_0(\cos k_x+\cos k_y+\cos k_z)$.
The 4$\times$4 pairing term is spin-singlet with equal orbital weights
$\hat\Delta(\bm r)=\Delta_0\sigma_0\otimes{\rm diag}\{f(\bm r-\bm r_1),f(\bm r-\bm r_2)\}$. Here, identity matrix $\sigma_0$ is in spin space, and the vortex function $f(\bm r)=\tanh(r/\xi)e^{i\theta}$ is defined for polar coordinate $\bm r=r(\cos\theta,\sin\theta)$ with coherence length $\xi$.

Then we can compute the topological invariant $\nu=\nu(\mu,\bm a)$ as a function of the chemical potential $\mu$ and the vortex core separation
$\bm a=\bm r_{1}-\bm r_{2}$.
The fourfold symmetry of the underlying square lattice implies the fourfold symmetry of $\nu$ with respect to $\hat{\bm a}\equiv\bm a/|\bm a|$. In the following we thus focus on the dependence of $\nu$ on $a\equiv|\bm a|$ and $\mu$, as shown in Fig.~\ref{fig3}.

When the bulk states and surface states are well separated by a global gap [Fig.~\ref{fig3}(a)], the corresponding phase diagram consists of two regions [Fig.~\ref{fig3}(b)]. 
When $\mu$ is inside the global gap, the topological phase is independent of the FV structure. 
When $\mu$ is outside the global gap, the bulk states introduce oscillatory patterns in the phase diagram related to the bulk Fermi momentum.

However, in iron-based superconductors, the surface states are masked by
bulk states at the same energy, as demonstrated by the band structures in Fig.~\ref{fig3}(c). As shown in the corresponding phase diagram in
Fig.~\ref{fig3}(d), the stable topological regions independent of the FV structure shrink, and MZMs can diminish when the FV-core separations
become too large. 

More details on the band structures and phase diagrams can be found in SM.

\begin{figure}
    \centering
    \includegraphics[width=1.0\linewidth]{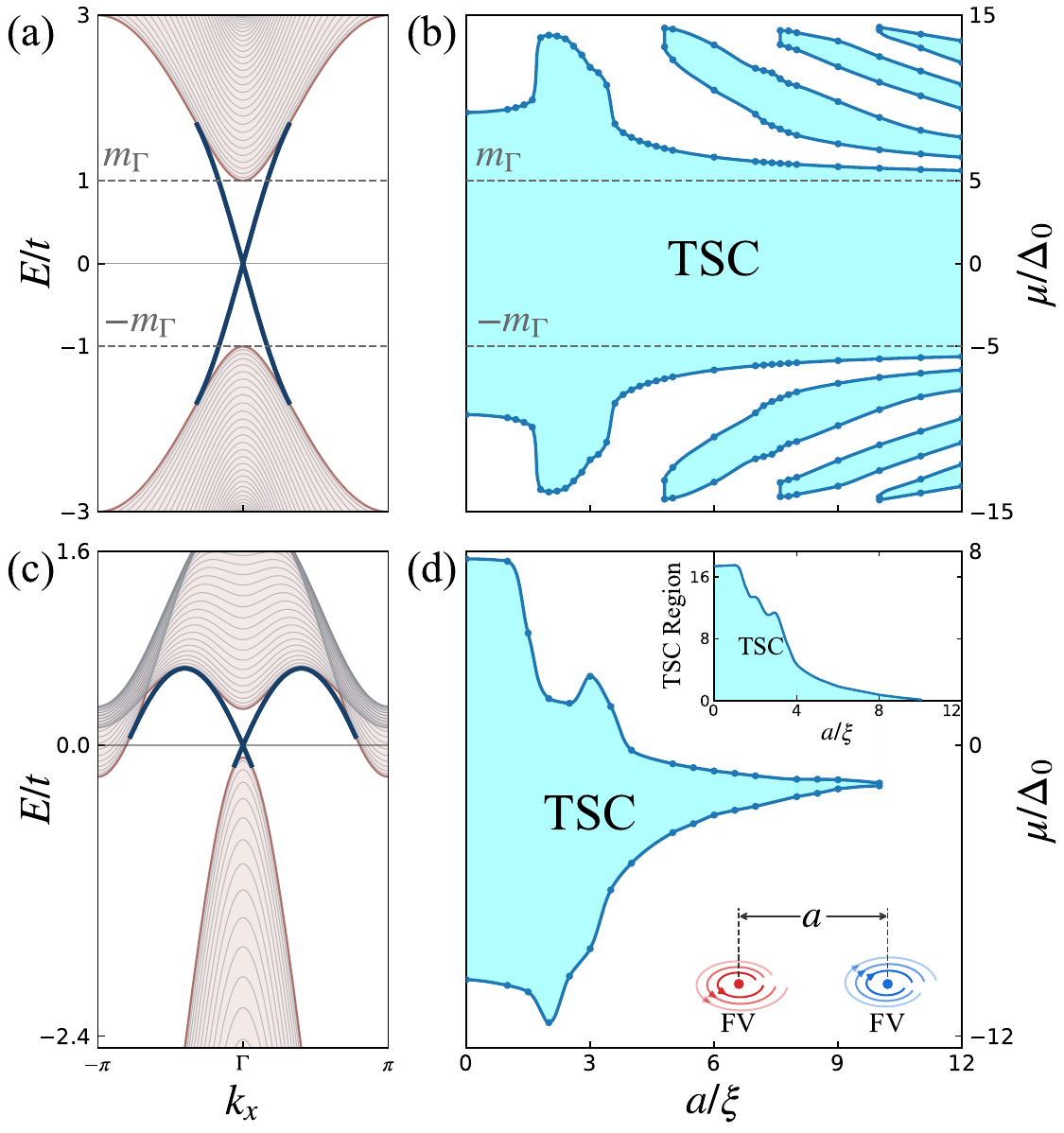}
    \caption{(a,c) Normal-state band structures and (b,d) superconducting-state phase diagrams of topological materials. In (a,c) the topological surface states (TSS) are colored in blue and bulk states red. In (b,d) two fractional vortices (FVs) are separated by distance $a$ with coherence length $\xi$ [(d) inset], and the chemical potential of the topological material is $\mu$ with pairing amplitude $\Delta_0$. In (a,b), the topological material has a global gap $m_{\Gamma}$. In (c,d), there is no global gap, and bulk states are involved in superconductivity for $\mu$ near the Dirac point. We used the lattice model with parameters \(t=0.5\), \(M=2.5\), \(m_0=-1.0\), and \(\Delta_0=0.1\), with \(q=0\) in (b) and \(q=-0.52\) in (d).}
    \label{fig3}
\end{figure}

\textit{\textcolor{blue}{Conclusion.}}---
In this work, we studied the properties of Majorana zero modes in fractional vortices which form \(\mathbb{C}P^{n-1}\) skyrmions \((n\geq 2)\), as a natural generalization of Fu-Kane mechanism in Abrikosov vortices.

Our results show that MZMs hosted by fractional vortices are spatially
unfolded over all vortex cores, with a size controllable by the FV-core
separations and the coherence length. This spatial unfolding not only
provides nonlocal experimental signatures such as crossed Andreev
reflection, but also offers multiple tunable knobs, including orbital
weights, core separations, and skyrmion number, for manipulating MZMs.
Beyond MZMs, other topological excitations, such as solitons and
fractionalized vortices themselves, may likewise serve as viable
candidates for topological quantum information processing and for
simulating exotic quantum matter. These possibilities position
fractional vortices and \(\mathbb{C}P^{n-1}\) skyrmions as a versatile
arena for exploring topological superconductivity and its applications.

Material systems capable of hosting both FVs and TSSs remain scarce. We propose CaKFe$_4$As$_4$ as a promising platform for realizing their coexistence. CaKFe$_4$As$_4$, whose crystal structure can be viewed as two different 122-type Fe-As superconductors CaFe$_2$As$_2$ and KFe$_2$As$_2$ inserted into one another, was the first iron-pnictide superconductor in which topological surface states and signatures of vortex-bounded MZMs were reported  \cite{Liu2020}. Recent experiments further revealed robust ZBCPs at surface steps in zero applied magnetic field, interpreted as possible MZM signatures associated with spontaneous time-reversal symmetry breaking  \cite{Fe2}. Within our theoretical framework, self-doping associated with distinct surface terminations could locally favor superconductivity with broken time-reversal symmetry, providing a route to the realization of MZMs in FVs. Spatially resolved scanning tunneling microscopy and spectroscopy could probe the associated vortex-core spectra and localized zero-energy states, providing an experimental avenue for testing our predictions.

\textit{\textcolor{blue}{Acknowledgment.}}---We thank Baiqing Lv, Qiangqiang Gu and Leo Kouwenhoven for inspiring discussions.
We acknowledge the support of NSFC grants No. 92565305, No. 12504164 and the Cultivation Project of Shanghai Research Center for Quantum Sciences (Grant No. LZPY2024).

\renewcommand{\thefigure}{S\arabic{figure}}
\renewcommand{\theequation}{S\arabic{equation}}
\renewcommand{\thetable}{S\arabic{table}}
\setcounter{figure}{0}
\setcounter{table}{0}
\setcounter{equation}{0}
\begin{widetext}
\title{Supplemental Material for “Majorana zero modes in fractional vortices"}

\maketitle

\section{Exact Solution of Surface Zero Mode}
\label{sec:surface}
We use the surface Nambu basis \((\psi_{\uparrow}^{\dagger},\psi_{\downarrow}^{\dagger},\psi_{\downarrow},-\psi_{\uparrow})\).  With
$\partial_\pm=\partial_x\pm\ii\partial_y$ and $\bm k=-\ii\bm\nabla$,
the surface Bogoliubov--de Gennes (BdG) operator is
\begin{equation}
 \mathcal H_s=
 \begin{pmatrix}
 -\mu & \vF\partial_- & \Ds &0\\
 -\vF\partial_+&-\mu&0&\Ds\\
 \Ds^*&0&\mu&-\vF\partial_-\\
 0&\Ds^*&\vF\partial_+&\mu
 \end{pmatrix}.
 \label{eq:surfaceBdG}
\end{equation}
Here the scalar $\Ds$ is the pairing in the time-reversed-hole
Nambu convention. 
For a particle--hole self-conjugate zero mode choose
$
 \Psi=(u,v,v^*,-u^*)^{\mathsf T}.
$

The first two rows of $\mathcal H_s\Psi=0$ give the equations:
\begin{equation}
 \vF\partial_+u+\Ds u^*+\mu v=0,
 \quad
 \vF\partial_-v+\Ds v^*-\mu u=0.
 \label{eq:coupled}
\end{equation}
The remaining rows are their complex conjugates.  This direct
substitution also fixes all signs used below.

Suppose that the surface pairing is a sum of phase-aligned, unit-winding
orbital contributions,
\begin{equation}
 \Ds(\bm r)=\sum_{\alpha=1}^{N}
 f_\alpha(\rho_\alpha)e^{\ii\theta_\alpha},
 \label{eq:pairingSum}
\end{equation}
where $\rho_\alpha=|\bm r-\bm R_\alpha|$,$
 \theta_\alpha=\arg(\bm r-\bm R_\alpha)$,$f_\alpha(0)=0$ and $f_\alpha$ is real.  Define the real function
\begin{equation}
 F_\alpha(\rho)=\int_0^\rho f_\alpha(s)\,\dd s,
 \quad W(\bm r)=\sum_\alpha F_\alpha(\rho_\alpha).
 \label{eq:W}
\end{equation}
Because $\partial_+\rho_\alpha=e^{\ii\theta_\alpha}$,
we have the equation$
 \Ds=\partial_+ W.
$
At $\mu=0$, Eqs.~\eqref{eq:coupled} therefore have the exact solution
\begin{equation}
 \displaystyle
 \Psi_0(\bm r)=\mathcal N e^{-W(\bm r)/\vF}
 (1,0,0,-1)^{\mathsf T}.
 \label{eq:exactzero}
\end{equation}
Indeed, $u=u^*$ and
$\vF\partial_+e^{-W/\vF}=-(\partial_+W)e^{-W/\vF}$,
while the second equation is identically zero.  If $\vF>0$ and
$W(\bm r)\to+\infty$ sufficiently rapidly, this solution is
normalizable with
\begin{equation}
 |\mathcal N|^{-2}=2\int_{\mathbb R^2}
 e^{-2W(\bm r)/\vF}\,\dd^2r.
 \label{eq:norm}
\end{equation}
Its exact density is $2|\mathcal N|^2e^{-2W/\vF}$, so its maxima
are precisely the global minima of $W$.  If
$f_\alpha(\rho)\to\Delta_\alpha^\infty>0$, then
$W=\sum_\alpha\Delta_\alpha^\infty\rho_\alpha+o(|\bm r|)$ far from
all cores.  The weighted Fermat--Torricelli construction therefore
describes the saturated-amplitude limit.  For a finite core profile,
the minimizer must instead be found from the full $W$ in
Eq.~\eqref{eq:W}; a flat minimizer segment need not be exact.

\begin{figure}
    \centering
    \includegraphics[width=1.0\linewidth]{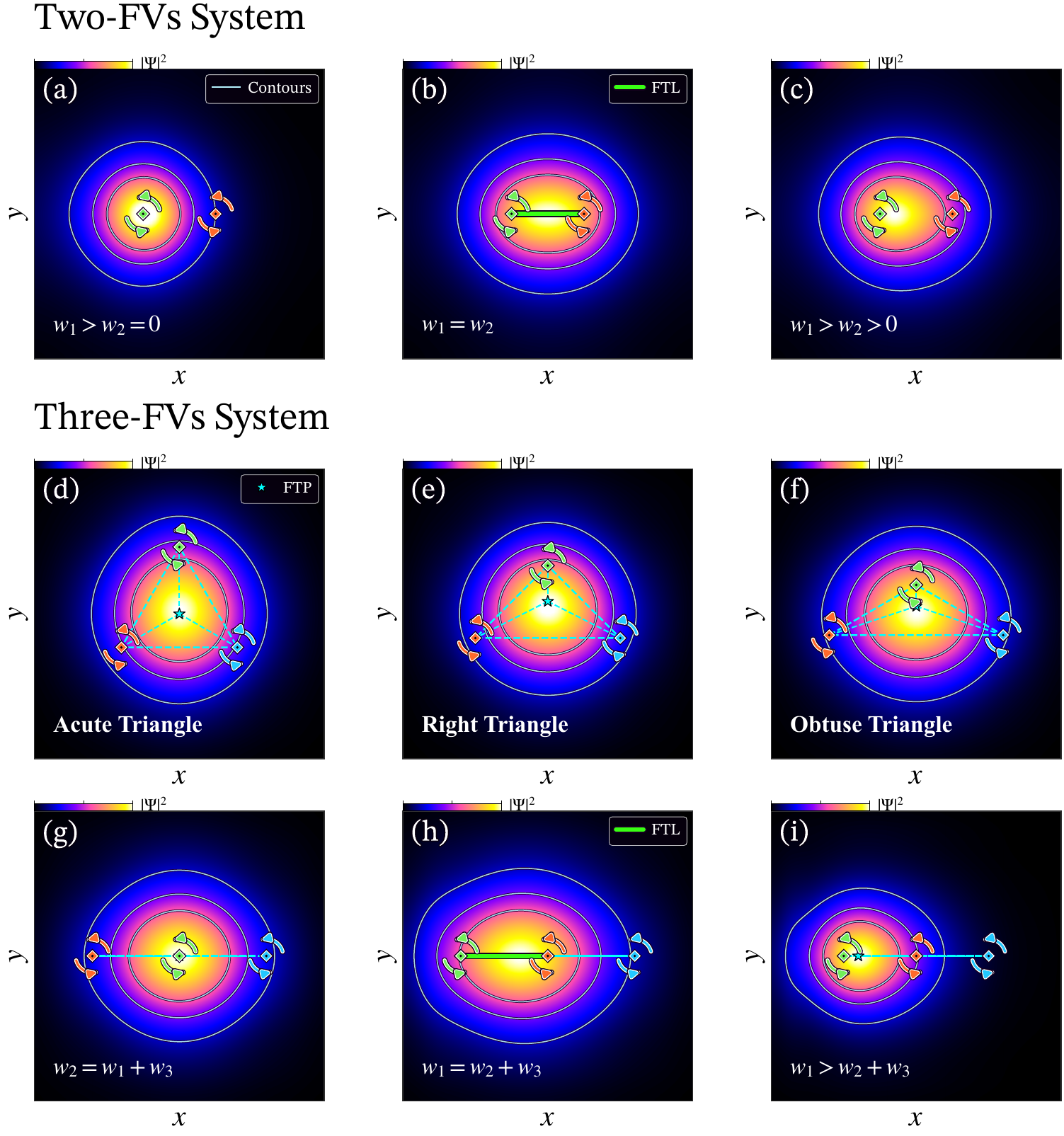}
    \caption{Spatial distributions of the surface zero modes in the two-FVs system and the three-FVs system under various orbital weight configurations.}
    \label{fig:mzm}
\end{figure}

The limiting weighted-distance function
$L(\bm r)=\sum_\alpha w_\alpha|\bm r-\bm R_\alpha|$
has simple exact minimizer criteria.  For two distinct cores,
$w_1>w_2$ gives the first core as the unique minimizer, while
$w_1=w_2$ gives the full inter-core segment; these statements
follow from the triangle inequality.  For three noncollinear
cores, a vertex $\bm R_i$ is a minimizer precisely when
\begin{equation}
 \left|\sum_{j\ne i}w_j
 \frac{\bm R_i-\bm R_j}{|\bm R_i-\bm R_j|}\right|\leq w_i.
 \label{eq:vertexcriterion}
\end{equation}
This is the subgradient condition at a cusp.  If no vertex
satisfies it, the unique interior minimizer obeys
$\sum_i w_i(\bm r-\bm R_i)/|\bm r-\bm R_i|=0$.
For collinear cores the minimizers are weighted medians, with
a whole interval only when the cumulative weights on its two
sides are equal.  These are claims about $L$; using them for the
finite-profile density requires checking the full $W$.

For comparison, a single radial vortex
$\Ds=f(r)e^{\ii\theta}$ has an exact finite-$\mu$ solution as well.
Set $W(r)=\int_0^r f(s)\,\dd s$ and
$u=e^{-W/\vF}A(r)$,
$v=e^{-W/\vF}e^{\ii\theta}B(r)$ with real $A,B$.  Then
\begin{equation}
 A'=-kB,\quad B'+\frac{B}{r}=kA,\quad
 k=\frac{\mu}{\vF}.
 \label{eq:singleODE}
\end{equation}
The solution regular at the origin is
\begin{equation}
 u=\mathcal N e^{-W/\vF}J_0(kr),\quad
 v=\mathcal N e^{-W/\vF}e^{\ii\theta}J_1(kr).
 \label{eq:singleBessel}
\end{equation}
The $B/r$ term, which is the transverse derivative of $v$ on a
radial line, is important in assessing a proposed one-dimensional
reduction for several vortices.

We numerically verified the systems with two fractional vortices and three fractional vortices, and the specific numerical method is given in the next section. By using different models and tuning the order parameters of each orbital, we calculated the spatial distribution of the surface zero modes in various cases, see in Fig. \ref{fig:mzm} and Table. \ref{tab:two_fvs_classification}\ref{tab:three_fvs_classification}, and verified that they all satisfy the laws of the exact solution.

\begin{table}[h]
\centering
\caption{Classification of the Two-FVs System}
\label{tab:two_fvs_classification}
\footnotesize
\setlength{\tabcolsep}{2pt}
\renewcommand{\arraystretch}{1.15}
\begin{tabular*}{\columnwidth}{@{\extracolsep{\fill}}
>{\raggedright\arraybackslash}p{\twoFVsWeightColumnWidth}
>{\raggedright\arraybackslash}p{\twoFVsContourColumnWidth}
>{\raggedright\arraybackslash}p{\twoFVsPeakColumnWidth}}
\hline\hline
\textbf{Weight Condition}
& \textbf{DOS Contours}
& \textbf{DOS Peak} \\
\hline
\ $w_m > w_n = 0$
& Concentric Circles
& Vortex Core \\
\ $w_m = w_n$
& Confocal Ellipses
& Vortex Line \\
\ $w_m > w_n > 0$
& 2-Focus Cartesian Ovals
& Vortex Core \\
\hline\hline
\end{tabular*}
\end{table}

\begin{table}[h]
\centering
\caption{Classification of the Three-FVs System}
\label{tab:three_fvs_classification}
\footnotesize
\setlength{\tabcolsep}{2pt}
\renewcommand{\arraystretch}{1.15}
\begin{tabular*}{\columnwidth}{@{\extracolsep{\fill}}
>{\raggedright\arraybackslash}p{\threeFVsWeightColumnWidth}
>{\raggedright\arraybackslash}p{\threeFVsGeometryColumnWidth}
>{\raggedright\arraybackslash}p{\threeFVsPeakColumnWidth}}
\hline\hline
\textbf{Weight Condition}
& \textbf{Geometric Configuration}
& \textbf{DOS Peak} \\
\hline
\multirow{3}{\threeFVsWeightColumnWidth}{\raggedright \ $w_i < w_j + w_k$ ($\forall i,j,k$)}
& Non-Collinear ($\forall i:\alpha_i < \theta_i$)
& Interior Point \\
& Non-Collinear ($\exists i:\alpha_i \geq \theta_i$)
& Vortex Core \\
& Collinear 
& Vortex Core \\
\hline
\multirow{3}{\threeFVsWeightColumnWidth}{\raggedright \ $w_i = w_j + w_k$}
& Non-Collinear
& Vortex Core \\
& Collinear (endpoint: $A_i$)
& Vortex Line \\
& Collinear (middle point: $A_i$)
& Vortex Core \\
\hline
\ $w_i > w_j + w_k$
& Arbitrary
& Vortex Core \\
\hline\hline
\end{tabular*}
\end{table}

Because it is difficult to construct a sufficiently large lattice in numerical calculations, one has to consider the influence of the order parameter within the coherence length of the vortex.
We now take the near-field limit into consideration. For a smooth superconducting pairing amplitude $f_\alpha(\rho) = \Delta_{s\alpha}^\infty \tanh(\rho/\xi)$, $W(\bm{r})$ possesses an exact solution:
\begin{equation}
W(\bm{r}) = \xi \sum_{\alpha=1}^N \Delta_{s\alpha}^\infty \ln\left( \cosh\left( \frac{|\bm{r} - \bm{r}_\alpha|}{\xi} \right) \right)
\end{equation}
In Fig. \ref{fig:mzm}, the FTP we label is the result after considering the near-field correction.

\section{Lattice models and numerical results}
\label{sec:num}

Numerically, we construct the 3D normal-state Hamiltonian on a simple cubic lattice and introduce s-wave superconductivity. On the Nambu basis $\Psi^\dagger_{\mathbf{k}}=(\bm c_{\uparrow}^{\dagger},\bm c_{\downarrow}^{\dagger},\bm c_{\uparrow},\bm c_{\downarrow})$, the full BdG Hamiltonian is
\begin{equation}
\mathcal{H}_{\mathrm{BdG}}(\mathbf{k},\mathbf{r})=
\begin{pmatrix}
H(\mathbf{k})-\mu & \Delta(\mathbf{r}) \\
\Delta^\dagger(\mathbf{r}) & \mu-H^{*}(-\mathbf{k})
\end{pmatrix},
\label{eq:threebdg}
\end{equation}
where $\bm c=\{c_{\sigma\alpha}\}$ is assumed to have spin $\sigma=\uparrow,\downarrow$ and $\alpha=1,\dots,N$ orbital components. $\mu$ is the chemical potential and $\bm k=-i\bm \nabla$ is the momentum operator. 
For each orbital, the pairing term is $\Delta_\alpha(\bm r)=\Delta_\alpha^{\infty}\tanh(|\mathbf{r}-\mathbf{r}_\alpha|/\xi)e^{i\theta}$ is defined for polar coordinate $\bm r=r(\cos\theta,\sin\theta)$ with coherence length $\xi$.

For the case where a vortex splits into two fractional vortices (FVs), we consider a system with two orbital degrees of freedom. Then the normal-state Hamiltonian can be expressed in terms of Dirac matrices as 
\begin{equation}
H^{(2)}(\bm k)=\varepsilon(\bm k)+\bm d(\bm k)\cdot\bm\Gamma+m(\bm k)\Gamma_0,
\end{equation}
with $\varepsilon(\bm k)=-2\sum_iq_i\cos k_i$, $d_i(\bm k)=2t_i\sin k_i$ and mass term $m(\bm k)=M+\sum_im_{0i}\cos k_i$, $i=x,y,z$.

The $\varepsilon(\bm k)$ term breaks the particle–hole symmetry (PHS) of the normal state and masks the surface states into the bulk states, which makes it closer to the properties of iron-based superconductors and will play a role in the calculation of the phase diagram.

When $\varepsilon(\bm k)=0$, the normal state preserves PHS and reduces to the Hamiltonian of a topological insulator. We refer to the 3D TI model proposed by Hosur et al.\ \cite{Hosur2011a, Hosur2010a} and generalize this model to the fractional-vortex case.

We construct BdG models realizing such configurations subject to two requirements:
(1).\ When the two vortices coincide, $\Delta_A(\mathbf{r})=\Delta_B(\mathbf{r})\equiv\Delta(\mathbf{r})$, these models should be unitarily equivalent to each other.
(2).\ The pairing matrix is diagonal in orbital space, i.e.\ intra-orbital rather than inter-orbital, so that the order parameter splits into two independent components living on the two orbitals, whose vortex centers $\mathbf{r}_A$ and $\mathbf{r}_B$ can be separated in real space.
(An inter-orbital pairing cannot be split in this way: fermion antisymmetry $\Delta^{T}=-\Delta$ relates its two off-diagonal blocks, so it carries only a single order parameter.) 

Insisting on a representation in terms of tensor products of Pauli matrices and imposing requirements (1) and (2) on the candidate models leaves exactly three inequivalent families. We list their normal-state Hamiltonians and pairing terms:

\emph{Family I.}
\begin{align}
H^{(2)}_{\mathrm I}(\mathbf{k})&=(d_x\sigma_x+d_y\sigma_y+d_z\sigma_z)\tau_x
+ m\tau_z  ,\\
\Delta_{\mathrm I}(\mathbf{r})&= i\sigma_y \otimes
\begin{pmatrix}\Delta_A(\mathbf{r})&0\\0&\Delta_B(\mathbf{r})\end{pmatrix}.
\end{align}

\emph{Family II.}
\begin{align}
H^{(2)}_{\mathrm{II}}(\mathbf{k})&=(d_x\sigma_x+d_y\sigma_y)\tau_z
+d_z\sigma_z \tau_z+m\tau_x  ,\\
\Delta_{\mathrm{II}}(\mathbf{r})&= i\sigma_y \otimes
\begin{pmatrix}\Delta_A(\mathbf{r})&0\\0&\Delta_B(\mathbf{r})\end{pmatrix}.
\end{align}

\emph{Family III.}
\begin{align}
H^{(2)}_{\mathrm{III}}(\mathbf{k})&=(d_x\sigma_x+d_y\sigma_y)\tau_z
+d_z\tau_x +m\tau_y  ,\\
\Delta_{\mathrm{III}}(\mathbf{r})&= i\sigma_y \otimes
\begin{pmatrix}\Delta_A(\mathbf{r})&0\\0&-\Delta_B(\mathbf{r})\end{pmatrix}.
\end{align}
where $\tau_i$ ($\sigma_i$) are Pauli matrices in orbital (spin) space.

Different families may introduce differences in the bulk states (e.g.\ , the bulk states of family I and family II differ, while those of family II and family III are the same), which will affect the vortex phase transition (VPT). 
Different families will also affect the orbital components of the topological surface states (TSSs), thereby changing the pairing felt by the TSS. For family III, the upper and lower surfaces each have only one orbital component, i.e.\ , the surface orbital weight of one of them is $w=0$. In this case, the MZMs at the two ends of the vortex will be spatially separated as the vortex splits into FVs.

For the case where a vortex splits into three FVs, we referred to the effective Hamiltonian of FST \cite{FeSCtheory2a} and simplified it to a certain extent, obtaining an model with four orbital degrees of freedom.
The 4-band effective model without SOC is restricted to the following form:
\begin{equation}
H^{(3)}_0(\mathbf{k})=
\begin{pmatrix}
E&0&0&0\\
0&m&0&i\kappa_-\\
0&0&m&i\kappa_+\\
0&-i\kappa_+&-i\kappa_-&-m
\end{pmatrix},
\end{equation}
where $\kappa_\pm=d_x\pm id_y$, $m=M+\sum_im_{0i}\cos k_i$.

Since the three-FVs system requires three order parameters, we decouple the orbital \(|1\rangle\) from the other orbitals in this model to save computational resources.
The $H_{\rm SOC}$ is given by
\begin{equation}
H_{\rm soc}^{(3)}(\mathbf k)=
\begin{pmatrix}
0&0&0&0&0&0&0&0\\
0&0&0&0&0&0&0&d_z\\
0&0&2\lambda&0&0&0&0&0\\
0&0&0&0&0&0&d_z&0\\
0&0&0&0&0&0&0&0\\
0&0&0&0&0&2\lambda&0&0\\
0&0&0&d_z&0&0&0&0\\
0&d_z&0&0&0&0&0&0
\end{pmatrix}.
\end{equation}

When calculating the spatial distribution of the surface zero modes, the model parameters we used are listed in Table. \ref{tab:parameters}.
\begin{table}[h]
    \centering
    \caption{Numerical Calculation Parameters}
    \label{tab:parameters}
    \setlength{\tabcolsep}{3pt}
    \renewcommand{\arraystretch}{1.25}
    \begin{tabular*}{\columnwidth}{@{\extracolsep{\fill}}*{4}{c}@{}}
        \hline\hline
        \multicolumn{4}{c}{Two-FVs System} \\
        \hline
        \(M\) & \((m_{0x},m_{0y},m_{0z})\) & \((t_x,t_y,t_z)\) & \((q_x,q_y,q_z)\) \\
        -2.0 & 1.0 & 0.4 & -0.52,-0.52,-0.45 \\
        \hline
    \end{tabular*}
    \par\nointerlineskip
    \begin{tabular*}{\columnwidth}{@{\extracolsep{\fill}}*{5}{c}@{}}
        \multicolumn{5}{c}{Three-FVs System} \\
        \hline
         \(M\) & \((m_{0x},m_{0y},m_{0z})\) & \((t_x,t_y,t_z)\) & \(E\) & \(\lambda\) \\
        -0.8 & -0.2,-0.2,0.4 & 0.075,0.075,0.21 & 3.0 & 1.2 \\
        \hline\hline
    \end{tabular*}
\end{table}

\section{Interpretation of the Phase diagram}
We found a phenomenological formula to understand the oscillatory topological phase. We consider that the period of the oscillation is related to the Fermi wave vector, and the Fermi wave vector \(k_F\) is related to the chemical potential \(\mu\),
\begin{equation}
    \mu^2=(M+3m_0-\frac{m_0}{2} k_F^2)^2+(2t)^2k_F^2.
\end{equation}
Thus we obtain an implicit equation involving $\mu$ and $a$,
\begin{equation}
\mu=J\sin\!\left[\sqrt{2}k_{\rm F}(a-a_c)+\Theta\right],
\end{equation}
where $J=1.4$, $a_c=1.8$ and $\Theta=0.7$.
In this formula, we conjecture that the oscillation takes the form of a sine wave, i.e., it does not decay as the vortex spacing increases. This is quite anomalous, yet it corresponds well with the numerical results of the phase diagram in Fig. \ref{fig:pd}.
The oscillation that does not decay with distance may be related to the coupling manner between fractional vortices. %There may exist a complex long-range interaction between fractional vortices, similar to the strong interaction between quarks.
\begin{figure}[t]
    \centering
    \includegraphics[width=0.95\linewidth]{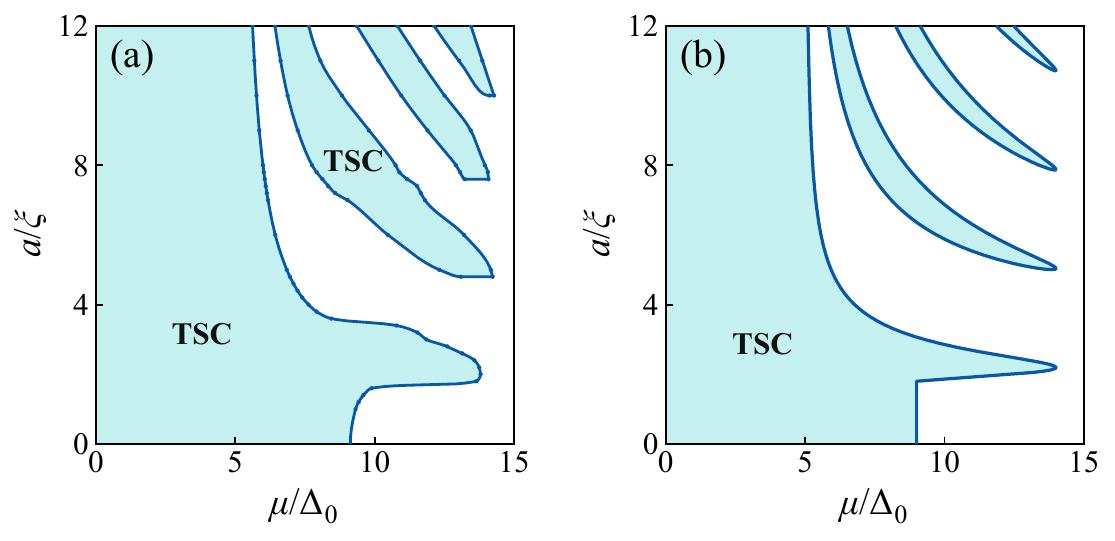}
    \caption{(a) is the phase diagram obtained from numerical calculations with parameters \(q=0\), \(t=0.5\), \(M=2.5\), \(m_0=-1.0\), and \(\Delta_0=0.1\). Due to the presence of PHS, only the \(\mu>0\) part is taken. (b) is the result fitted by the theoretical formula.}
    \label{fig:pd}
\end{figure}

\end{widetext}

\end{document}